# Electron-nuclear coherent coupling and nuclear spin readout through optically polarized $V_B^-$ spin states in hBN

F.F. Murzakhanov,[1] G.V. Mamin,[1] S.B. Orlinskii,[1] U. Gerstmann,[2] W.G. Schmidt,[2] T. Biktagirov,[2,*] I. Aharonovich,[3] A. Gottscholl,[4] A. Sperlich,[4] V. Dyakonov,[4] and V.A. Soltamov[1,†]

[1]*Kazan Federal University, Kazan, 420008 Russia*
[2]*Theoretische Materialphysik, Universität Paderborn, 33098 Paderborn, Germany*
[3]*School of Mathematical and Physical Sciences, University of Technology Sydney, Ultimo, NSW, Australia*
[1]*Experimental Physics 6, Julius Maximilian University of Würzburg, 97074 Würzburg, Germany*

* timur.biktagirov@upb.de
† victrosoltamov@gmail.com

**ABSTRACT**: Coherent coupling of defect spins with surrounding nuclei along with the endowment to read out the latter, are basic requirements for an application in quantum technologies. We show that negatively charged boron vacancies ($V_B^-$) in electron-irradiated hexagonal boron nitride (hBN) meet these prerequisites. We demonstrate Hahn-echo coherence of the $V_B^-$ electron spin with a characteristic decay time $T_{coh}$ = 15 µs, close to the theoretically predicted limit of 18 µs for spin defects in hBN. Modulation in the MHz range superimposed on the Hahn-echo decay curve are shown to be induced by coherent coupling of the $V_B^-$ spin with the three nearest $^{14}$N nuclei through a nuclear quadrupole interaction of 2.11 MHz. Supporting DFT calculation confirm that the electron-nuclear coupling is confined to the defective layer. Our findings allow an in-depth understanding of the electron-nuclear interactions of the $V_B^-$ defect in hBN and demonstrate its strong potential in quantum technologies.

**KEYWORDS:** *hexagonal boron nitride, boron vacancy, optical spin polarization, electron-nuclear coherent coupling*

Optically addressable high spin states ($S \geq 1$) of defects in a semiconducting host are fundamental for the development of solid-state quantum technologies and serves as a tool to probe a plethora of novel physical phenomena [1, 2]. Nanoscale quantum sensing [3], quantum informational processing[1, 4] and realization of exotic states of matter [2] are just a few examples. Mainly two solid-state platforms have been intensively explored within this respect. Namely, diamond with negatively charged nitrogen-vacancy centers (NV⁻) and silicon carbide (SiC) comprising several silicon vacancy-related defects [1, 5–7]. Both 3D crystals are formed by $sp^3$-hybridized atoms of nearly non-magnetic (i.e. low-abundant magnetic) nuclei. They are, thus, rather similar with respect to their structural and chemical properties.

A completely different class of host materials for defect spins has been recognized only recently by referring to the van der Waals' materials family, such as hexagolnal Boron Nitride (hBN), after successful demonstration of optical and microwave control over the spins bounded to defects in its lattice by means of optically detected magnetic resonance (ODMR) [8–11]. hBN is formed by 2D atomic layers of $sp^2$-hybridized Nitrogen-Boron atoms that are coupled through weak vdW interactions. Its ultrawide-bandgap ($E_g \approx 6$ eV) [12] combined with two-dimensional screening mechanisms, and natural hyperbolic properties in the mid-infrared range, make such kind of layered material particularly interesting for quantum technologies [13]. Defects in hBN are deeply explored from the perspective of quantum photonics [8–11, 13, 16, 17], however, so far only one defect possessing ODMR has been rigorously identified and its microscopic structure is well understood. This defect is the negatively-charged boron vacancy ($V_B^-$), a missing boron atom having three equivalent nitrogen atoms as nearest neighbors, as schematically shown in Fig. 1 (a) [8, 16, 17]. $V_B^-$ possesses a spin-triplet (*S*=1) ground state, whereby the spin-spin interaction between the unpaired spins induces energy-level splittings (*D* ≅ 3.6 GHz) even in the zero magnetic field. A spin-dependent recombination channel in its optical excitation-recombination cycle (Fig. 1(b)) allows to polarize the $m_S = 0$ ground state sublevel and to realize its readout via ODMR or electron spin resonance (ESR). Taking into account (*i*) the demonstrated integration of $V_B^-$ defects into advanced photonic cavities [18], (*ii*) the high readout contrast of its ODMR signal, comparable with those of NV⁻ centers in diamond [19], and (*iii*) the developed robust mechanisms of defect generation in hBN [8, 20–22], $V_B^-$ centers provide a promising platform to probe and extend the quantum technologies concepts on defects spins confined in a two-dimensional atomic layer. However, both N and B are 100%-magnetic nuclei rendering a detailed understanding of the electron-nuclear coupling as a prerequisite for spintronic application.

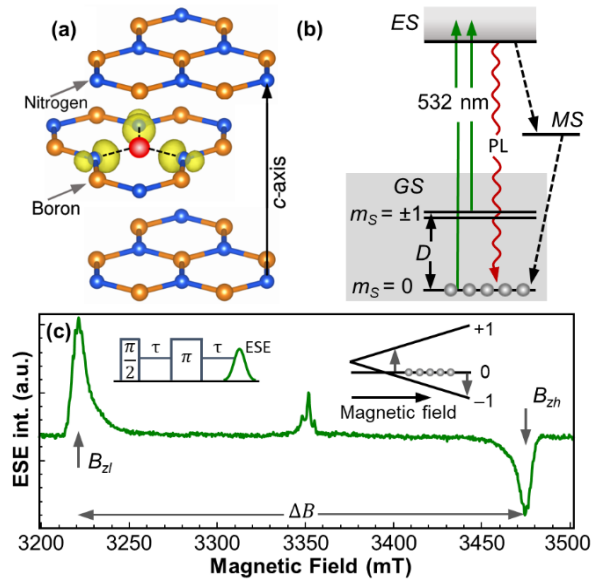

**Figure 1.** (a) hBN lattice with a $V_B^-$ defect (red ball). The *c*-axis is shown with the solid arrow. Nitrogen dangling bonds are shown with dashed lines. (b) Scheme of the $V_B^-$ spin initialization through spin-dependent recombination pathway (dashed lines) from the excited state (ES) into the ground state (GS, with occupied $m_S$ = 0 sublevels and ZFS denoted by *D*) via metastable state (MS) under *λ* = 532 nm excitation (green lines) and radiative decay (PL) from the ES to the GS (purple curve). (c) ESR spectrum of the $V_B^-$ defects at a temperature *T* = 50 K under *λ* = 532 nm excitation, obtained using Hahn-echo microwave pulse sequence (top inset) with static magnetic field $B_0 \| c$. Vertical arrows indicate allowed ESR transitions ($B_{zl}$ and $B_{zh}$) separated by Δ*B*.

In this article, we study the coherent coupling of optically polarized $V_B^-$ electron spin with surrounding nuclei by tracking the oscillation behavior of Hahn-echo coherence followed by an analysis of the coherence behavior through density functional theory (DFT). Understanding the interactions allowed us to implement a standard double resonance technique (ENDOR) to demonstrate the possibility to readout the corresponding nuclear spin. As a result, we are able to determine the complete ground state spin-Hamiltonian of the $V_B^-$ center and to perform the basic demonstration of the defect's prospective for quantum technologies.

The sample used in this study was a commercially produced (hq Graphene company) hBN single crystal which has been irradiated at room temperature with 2-MeV electrons to a total dose of $6 \times 10^{18}$ cm⁻² in order to produce the $V_B^-$ defects [21]. All experiments were done in the W-band (≅ 94 GHz) commercial pulsed ESR spectrometer Bruker Elexsys 680.

First, we measure the field-swept Electron Spin-Echo (ESE) detected ESR spectrum of the $V_B^-$ defects using Hahn-echo pulse sequence (inset in Fig. 1(c)). Two pronounced ESR lines in the magnetic fields labeled $B_{zl}$ and $B_{zh}$ correspond to the allowed magnetic dipole transitions between the triplet spin manifold (see also right inset in Fig. 1(c)). The splitting between these lines is $\Delta B \cong 255$ mT $= 2D/\gamma_e$, where $D = 3.57$ GHz and $\gamma_e = 28$ GHz/T is the electron gyromagnetic ratio, corresponding to the spectroscopic signatures of the $S = 1$ $V_B^-$ centers [8]. The spectrum also demonstrates that 532 nm excitation initialize the $m_S = 0$ ground-state spin sublevel of the defect through spin-dependent intersystem-crossing (as illustrated in Fig. 1(b)). The latter follows from phase reversal of the ESR signals, i.e. the $B_{zh}$ transition ($m_S = 0 \rightarrow -1$) exhibits emission rather than absorption of the microwave power.

We then probe the coherence of the optically polarized $V_B^-$ spin ensemble measuring a decay of the Hahn-echo magnitude by incrementing the delay time $\tau$ after the first microwave $\pi/2$-pulse (Fig. 2(a)). The decay curve measured on the $B_{zl}$ resonance is well described by a stretched exponent of the form $I(2\tau) = I_0 * exp\{-(2\tau/T_{coh})^n\}$ with $T_{coh} = 15.1$ μs and a moderate stretching parameter $n = 1.4$. From this, two important conclusions follow: (*i*) The measured coherence time $T_{coh}$ in this electron-irradiated sample is approximately 7 times longer than the Hahn-echo coherence of about 2 μs previously measured on the neutron-irradiated hBN sample [10, 23] and is longer than 10 μs – the upper bound of $V_B^-$ spin ensembles coherence estimated from the Rabi nutations [8]. Remarkably, this coherence is close to the 18 μs upper limit predicted for spin defects in hBN using a cluster expansion method [24]. It is thus within the common range of the optically polarized defect spins in diamond [25, 26] and SiC [27, 28]. (*ii*) The observed stretched exponential decay indicates that the total coherence time, in addition to spin-spin relaxation, is determined by spectral diffusion which is commonly characterized by a typical value of *n* between 1 and 4 depending on the regime of spectral diffusion [29]. The latter indicates coherent interaction of the $V_B^-$ defect spin with the temporally fluctuating random local (effective) magnetic field associated with the dipolar-interaction induced flip-flops of nuclear spin pairs [30] of $^{10}B$, $^{11}B$, and $^{14}N$.

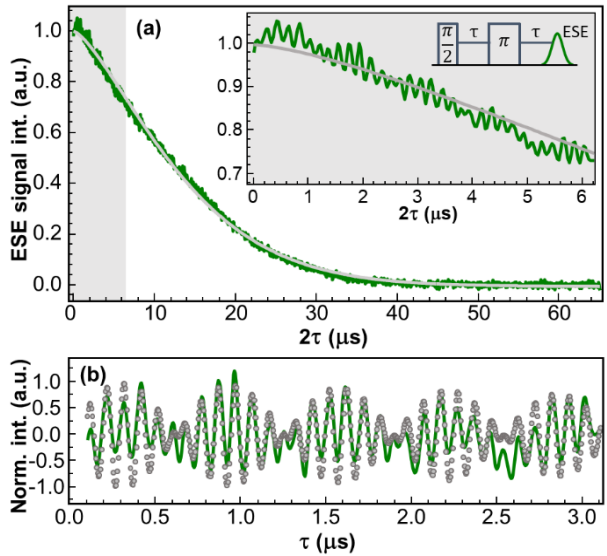

**Figure 2.** (a) Hahn-echo ESE decay curve (green) measured on the optically polarized $V_B^-$ spins as a function of the time delay $\tau$ between the mw pulses. Stretched exponential fit (gray line) demonstrates a characteristic decay time $T_{coh} = 15.1$ μs. Inset shows the gray-shaded segment in enlarged scale. (b) Experimentally observed modulation (green line) after subtraction of the corresponding decay curve; fit functions (gray circles, see text).

The decay curve reveals its oscillatory behavior especially pronounced at the very beginning of the transient curve (inset to Fig. 2(a)). Such oscillations refer to Electron Spin Echo Envelop Modulation (ESEEM) [32, 33] and manifest the presence of coherent coupling of the $V_B^-$ electron spin with magnetic moments of nuclei available in the hBN lattice. The modulation is generated by high-intense microwave pulses inducing mixing of the "forbidden" and "allowed" ESR transitions. As a result, the decay curve is modulated by frequencies corresponding to nuclear magnetic resonance (NMR) transitions within nuclear-spin manifolds bound to the current electron spin, whereby both sublattices, N as well as B, are potentially contributing.

To further elucidate the origin of these modulations we first analyze them in the time-domain after removal of the exponential contribution. The corresponding fit $f(\tau) = A_0 * cos(2\pi\tau f_1 + \varphi_1) * cos(2\pi\tau f_2 + \varphi_2)$ presented in Fig. 2(b) shows that the modulation pattern is described well by a product of a slow-oscillating envelope function ($f_2$ = 0.789 ± 0.001 MHz) and a fast-oscillating one ($f_1$ = 9.99 ± 0.09 MHz), whereby the latter (i.e. the average of contributing NMR frequencies) coincides with the Larmor frequency of $^{14}$N nuclear magnetic moments ($\nu_L(^{14}N)$) for the given magnetic field via $\nu_L = \gamma_n B_{zl} \approx$ 9.92 MHz, where $\gamma_n$= 3.077 MHz/T [31] is the $^{14}$N nuclear gyromagnetic ratio. The observed modulation can be thus naturally explained by coherent coupling of the $V_B^-$ electron spin with nuclear spins of the nitrogen sublattice.

To identify the particular $^{14}$N atoms giving rise to this coupling, and in order to establish the interaction type we determine the parameters of the electron-nuclear coupling and compare them with theoretical preditions from DFT modelling. For this purpose, we consider the following spin-Hamiltonian $H = \gamma_e B_0 S_z + D\left(S_z^2 - \frac{1}{3}S(S+1)\right) + SAI - \gamma_n I_z B_0 + IPI$ with electron Zeeman ($Z_e$), ZFS, hyperfine (HF), nuclear Zeeman ($Z_N$), and quadrupole (QI) interactions, where A and P are the HF and QI tensors, respectively. A static magnetic field $B_0$ is applied parallel to the c-axis which is the principle z axis of the axially-symmetric ZFS tensor, expressed by the D-value. For simplicity, we consider here a single interaction energy for the three nearest nitrogen nuclei exclusively. Due to the trigonal symmetry of the given ground state configuration these nuclei are indeed equivalent, and the respective nitrogen dangling bonds are in the given defective 2D B-N $sp^2$ plane (see also defect model in Fig. 1(a)). This implies that the symmetry axises of the HF and QI interactions are perpendicular to the c-axis and possess axial symmetry along the bonds, i.e. HF and QI interactions are fully described as $A = a + b(3cos^2\theta - 1)$ and $QI = P(3cos^2\theta - 1)$, respectively. Here $a$ and $b$ are isotropic and anisotropic part of the HF, respectively, and $P = \frac{3eQ_N V_{zz}}{4I(2I-1)}$ is related to the electric field gradient $V_{zz}$ in the direction of the nitrogen dangling bond and the nuclear electric quadrupole moment $Q_N$. $\theta$ is the angle between $B_0$ direction and the symmetry axis of the HF and QI and thus $\theta$ = 90°. The corresponding NMR resonance conditions for nuclear spin flips available in such system and thus potentially expected in the echo envelope are labeled in Fig. 3(b) as $\nu_1$ to $\nu_4$. These frequencies are $\nu_{1,2} = \nu_L \pm \frac{P}{2}$ and $\nu_{3,4} = A - \nu_L \pm \frac{P}{2}$, where the absolute value of $A$ in $B_0||c$ configuration has been established previously to be 47 MHz [8]. It is seen, that $\nu_3$ and $\nu_4$ correspond to the nuclear spin-flips between hyperfine- and quadrupole-splitted $m_S$ = +1 Zeeman level, while $\nu_1$ and $\nu_2$ frequencies couple purely quadrupole-splitted nuclear states in the $m_S$ = 0 sublevel. The latter two frequencies are symmetrically placed around the $\nu_L(^{14}N)$ frequency and are thus in accordance with the experimentally observed modulation.

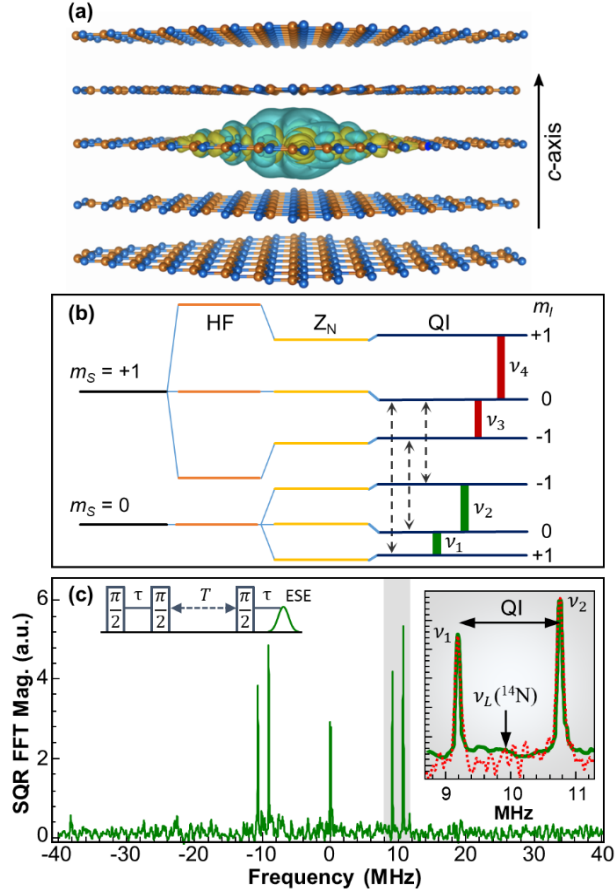

**Figure 3.** (a) Electron-spin density around the $V_B$ defect confined to the respective defect-containing BN layer. (b) Energy-level scheme of the $V_B^-$ ground state including $Z_e$, ZFS, HF, $Z_N$, and QI interactions with one $^{14}$N nucleus. Electron and nuclear levels are indexed by corresponding quantum numbers $m_S$ and $m_I$. NMR transitions $\nu_1 - \nu_4$ are denoted by green and red vertical bars. (c) FT spectrum of the three-pulse ESEEM, obtained using the pulse sequence (left inset). Right inset shows FT frequencies in enlarged scale together with the FT spectrum (red dots) of the modulations from Fig. 2(b); $\nu_L(^{14}N)$ denoted with a vertical arrow.

To analyse these $\nu_{1,2}$ frequencies precisely, we conduct three-pulse ESEEM experiments. Compared with the 2-pulse technique, a three-pulse method allows to reveal the basic frequencies of nuclear-spin flips, and eliminates observation of the frequencies given as sums and differences of the basic NMR transitions. A Fourier-Transform (FT) spectrum of such a measured echo modulation is presented in Fig. 3(c). The resonance frequencies labeled $\nu_1$ and $\nu_2$ are observed exactly as expected for the QI-induced splitting of the $m_S = 0$ sublevel. The value of the QI determined as the difference between the frequencies $|\nu_1 - \nu_2| = 2f_2$ is 1.58 MHz corresponding to the nuclear quadrupole interaction constant defined as $C_q = eQ_N V_{zz} = 2.11$ MHz. Notably this value is around 10 times larger than recently determined $C_q \cong 200$ kHz for the electric quadrupole moment of $^{14}$N nuclei interacting with the native electric field gradient of hBN [34]. On the other hand it is close to previously determined $C_q$ for the threefold coordinated nitrogen nucleus involved in the prototypical nitrogen-vacancy (NV$^-$) spin-qubits in diamond [35] and SiC [27].

We further support this finding by performing *ab initio* calculations within the framework of periodic density functional theory (DFT). A single $V_B^-$ center is embedded in a hBN supercell containing 800 atoms ($10 \times 10 \times 2$ unit cells with 4 BN layers). The calculations utilize the GIPAW module of the Quantum ESPRESSO software [37, 38], Perdew-Burke-Ernzerhof (PBE) exchange-correlation functional [39], standard norm-conserving pseudopotentials [40], and a plane-wave basis set with 600 eV kinetic energy cutoff. Our calculation confirm that the ground state provides trigonal symmetry, whereby the spin density is perfectly confined in the defective BN-layer, as shown in Fig.3 (a). Due to the symmetry of the charge density distribution in the vicinity of the $V_B^-$ center, the leading component of the electric field gradient at the nearest-neighbor nitrogen nucleus, $V_{zz}$, coincides with the direction of the nitrogen dangling bond. The value of $C_q$ = 2.06 ± 0.08 MHz derived from $V_{zz}$ shows very good agreement with experiment (2.12 MHz). The uncertainty in the calculated $C_q$ reflects the spread of $Q_N$ between 0.0193 and 0.0208 barn reported in literature [41]. We find also nice agreement with the other characteristic fingerprints of the center provided in the caption of Fig. 4: the ZFS and the out-of-plane HF (for $\theta$ = 90°) are calculated to be $D$ = 3.47 GHz and $A$ = 46.12 MHz, respectively. As for the other $^{14}$N nuclei in the supercell, the calculated NQI is about five times weaker with $C_q$ of 0.35 ± 0.01 MHz for the next-nearest-neighbour nitrogens and 0.42 ± 0.01 MHz for the most distant ones.

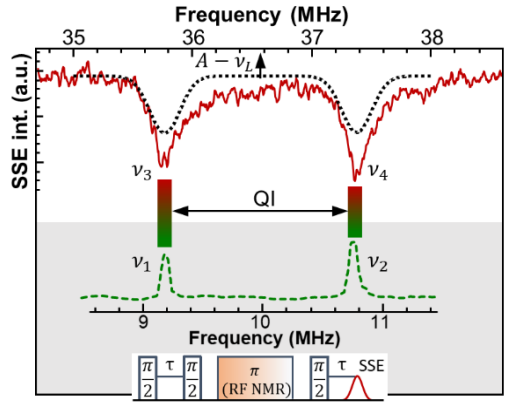

**Figure 4.** Readout of the $\nu_3$ and $\nu_4$ NMR transitions between quadrupole-splitted hyperfine levels in the $m_S$ = +1 state by monitoring the SSE intensity. Pulse sequence used for nuclear-spin readout indicated in the bottom. Calculated spectrum using the spin-Hamiltonian with parameters $D$ = 3.57 GHz, $A$ = 46.52 MHz, $C_q$ = 2.12 MHz is shown with black dashed line. Bottom spectrum reflects the QI in the $m_S$ = 0 state, taken from the inset of Fig. 3(c).

The intricate peculiarity of interaction with more distant $^{14}$N shells can be eliminated by direct manifestation of the 1.58 MHz interaction in the hyperfine-splitted $m_S$ = + 1 state as the latter originates from the three equivalent nitrogen atoms nearest to the vacancy. In Fig. 4 this is demonstrated by probing the population of the hyperfine sublevels through the generation of the stimulated electron spin echo (SSE) following the driving of appropriate nuclear magnetic resonance transitions. The pulse sequence used in this experiment (Mims-ENDOR [42]) is shown in the inset of Fig. 4. The first two π/2 mw pulses invert the electron spin population; the third π/2 pulse generates the stimulated ESE (SSE) signal. Between the second and the third mw pulses, a radiofrequency (RF) pulse is applied to invert the population of the nuclear spin sublevels, i.e. inducing NMR transitions. The resulting spectrum of $B_{zI}$ transitions is presented in Fig. 4. It possesses the doublet of lines labeled as $\nu_3$ and $\nu_4$ in accordance with the energy level diagram for the $m_S$ = +1 state. The center of gravity of these frequencies is on the value $A - \nu_L$. Their splitting $|\nu_3 - \nu_4|$ mimicks again the 1.58 MHz QI modulating the FT ESEEM spectrum of the $m_S$ = 0 state, cf. gray-shaded bottom in Fig. 4. This reinforces our aforementioned conclusion that the modulating 1.58 MHz are induced by $^{14}$N quadrupole interaction from the three nitrogen dangling bond atoms.

To conclude, in this letter we perform an in-depth analysis of the $V_B^-$ spin ensembles in hBN. We demonstrate their coherent coupling with atoms of the nitrogen sublattice, identify the origin of this coupling and unambiguously reveal it to be induced by nuclear quadrupole interaction NQI of the three neighboring $^{14}$N dangling bond atoms. We find that the magnitude of this interaction is about one order larger than that previously established for the host nitrogen atoms. The 15 $\mu$s spin-coherence time measured here is apparently longer than previously reported for the $V_B^-$ spin ensemble in neutron-irradiated samples and is thus within the range commonly observed on optically polarized spin ensembles of defects in diamond and SiC. Additionally we reveal that the coherence of the $V_B^-$ defect possess nontrivial stretched-exponential behavior pointing to the presence of coherent interactions of the $V_B^-$ defect spin with fluctuating spin bath generated by a distant nuclei. These all together make the $V_B^-$ defect spin confined in the two-dimensional boron nitride layer an intriguing object for implementations in quantum technologies and potentially in the discovery of the novel quantum states in the limits of condensed matter.

## ACKNOWLEDGMENTS

This work has been supported by the RSF grant No. 20-72-10068. Numerical calculations were performed using grants of computer time from the Paderborn Center for Parallel Computing (PC$^2$) and the HLRS Stuttgart.